\documentclass[12pt]{JHEP3}

\usepackage{amsmath,amssymb,amsfonts}
\usepackage[dvips]{graphicx}

\newcommand{\be}{\begin{equation}}
\newcommand{\ee}{\end{equation}}
\newcommand{\ba}{\begin{eqnarray}}
\newcommand{\ea}{\end{eqnarray}}
\newcommand{\bea}{\begin{array}}
\newcommand{\eea}{\end{array}}

\newcommand{\Tr}{{\rm Tr}}

\makeatletter \@addtoreset{equation}{section} \makeatother

\preprint{KIAS-P08017}

\title{\Large \bf Janus and  Multifaced  Supersymmetric   Theories}

\author{Chanju Kim \\
Department of  Physics \\
Ewha Womans University \\
Seoul 120-750, Korea \\
\email{cjkim@ewha.ac.kr}  } 

\author{ Eunkyung Koh \\
Department  of Physics \\
Seoul National University \\
Seoul 130-743, Korea \\
\email{ekoh@phya.snu.ac.kr}  }

\author{Ki-Myeong Lee \\
School of Physics \\
Korea Institute for Advanced Study \\
Seoul 130-012, KOREA\\
\email{klee@kias.re.kr} }

\abstract{We investigate the various properties  of Janus supersymmetric
  Yang-Mills theories. A novel vacuum structure is found and  BPS
  monopoles and dyons are studied. Less supersymmetric Janus theories
  found before   are derived by a simpler method. In addition,  we
  find  the supersymmetric theories   when the coupling  constant
  depends on two and three spatial coordinates.  }

\begin{document}

\section{Introduction and Conclusion}

The AdS-CFT correspondence gives rise to many insights to the
conformal field theories~\cite{Maldacena:1997re,Gubser:1998bc,Witten:1998qj}.
The most successful example is the relation between the string
theory on $AdS_5\times S^5$ and the 4-dim ${\cal N}=4$
supersymmetric Yang-Mills theories. The original Janus solution in
Ref.~\cite{Bak:2003jk} is a 1-parameter family of dilatonic
deformations of $AdS_5$ space without supersymmetry. This solution
turns out to be stable under a large class of
perturbations~\cite{Bak:2003jk,Freedman:2003ax,Celi:2004st} and some
holographic properties have been explored  in
Refs.~\cite{Bak:2003jk,Freedman:2003ax,Papadimitriou:2004rz} The
Janus solution is made of  two Minkowski spaces joined along an
interface so that the dilaton field interpolates two asymptotic
values. The CFT dual field theory is suggested to be
the deformation of the  Yang-Mills theory where  the coupling
constant changes from one region to another
region at 2-dim interface~\cite{Bak:2003jk,Clark:2004sb}.

Further works revealed that one can have  supersymmetric Janus
geometries with the various supersymmetris and internal
symmetries\cite{Clark:2004sb,Clark:2005te,D'Hoker:2006uu}. Starting from the 16
supersymmetric Yang-Mills theory, the various deformations of 0,
2, 4, 8  supersymmetries  have  been found\cite{D'Hoker:2006uv}.
 Especially, the 16 supersymmetric Janus geometries
have been found~\cite{Gomis:2006cu,D'Hoker:2007xy,D'Hoker:2007xz}. Also  other aspects of the
Janus solutions have been discussed in
Ref.~\cite{Bak:2004yf,Sonner:2005sj,Bak:2006nh,Hirano:2006as,Bak:2007jm}.

Instead of following the detail of the frame work given in
Ref.~\cite{D'Hoker:2006uv} where the 6-dim symplectic Majorana
fermions are used extensively, we start from the 10-dim supersymmetric
Yang-Mills theory where the discussions are quite simple. In this
work, we give  a simple derivation of the deformation of the 16
supersymmetric Yang-Mills theory.

One could ask whether there is a supersymmetric deformation of the
Yang-Mills theory where the coupling constant depends on time too.
Indeed there have been several works along this direction~\cite{Chu:2006pa,Lin:2006ie,Das:2006dz,Chu:2007um}.
To maintain some supersymmetry, the time dependency of the coupling constant
should accompany the spatial dependency, say  $e^{2}(t+x)$.  It
turns out that there is no need to correct the Lagrangian or the
supersymmetric transformation besides reducing the supersymmetry by 1/2 by
imposing a constraint on the supersymmetry parameter spinor.

Starting from 10-dim supersymmetric Yang-Mills theories, one
may wonder about the higher dimensional Janus theories. For the simplest case
with 8 supersymmetries, one can easily read off from the Lagrangian that
such theory can exist in 7-dim spacetime as one needs 3 scalar fields. 
For the less supersymmetric case one needs more scalar fields, and 
so lower dimension. Additional spatial dependency of the coupling constant also needs
more scalar fields to maintain some supersymmetry.  Results in the Sec.6 and Sec.7 
casshows  the maximum spacetime dimension, depending on the cases.

The supersymmetric vacuum of the 8 supersymmetric Janus is governed by the Nahm equation~\cite{Nahm:1979yw}.
Besides the usual Coulomb phase, there can be nontrivial vacuum where the nonabelian 
gauge symmetry is completely broken near the planes where the coupling constant $e^2(z)$ can vanish.
In addition, one can have 1/2 BPS magnetic monopoles and charged particles and 1/4 BPS
dyons in the Coulomb phase.

In the limit of a sharp interface, one needs various continuity condition on the fields.
Especially one can see that there are mirror charges for magnetic monopoles and
electrically charged particles in the Coulomb phase. An incident massless wave on
a sharp face are partially reflected and partially transmitted without refraction.

In this work we study in detail the  properties of 8
supersymmetric Janus Yang-Mills theories, like the vacuum
structure and the BPS configurations. In addition, we recapitulate
the less supersymmetric Janus theories found in~\cite{D'Hoker:2006uv}. Then we
classify all the supersymmetric deformations of the 16 
supersymmetric Yang-Mills theories when the coupling
constant depends on the two or three spatial coordinates. 
These
higher dimensional cases tend to have less supersymmetries.
We have not explored in the detail the properties of these less supersymmetric
theories. There may be some surprises.  Nonsupersymmetric
geometry with a special higher dimensional Janus type has been worked out~\cite{Hirano:2006as}.
Our work suggests a possibility of supersymmetric Janus geometries where the dilaton field depends on
several coordinates.

When one has a theta term which also depends on the coupling constant, 
one may wonder there can be a supersymmetric theory. For example, 
the Yang-Mills parts of the Lagrangian can be written as
\be \Tr \frac{1}{4e^{2} } 
\left( -F_{\mu\nu} F^{\mu\nu} + \tan\alpha F_{\mu\nu}\tilde{F}^{\mu\nu} \right) \ , 
\ee
where $\tan\alpha= e^{2}\theta/8\pi^{2}$. As one can obtain the Janus geometry where both
dilaton and axion changes by the $SL(2,R)$ transformation~\cite{D'Hoker:2007xy}, one expects
a supersymmetric Lagrangian with the theta-term. However, we have not found
one yet.

Our analysis of Janus theories are done in the classical level. Once the quantum effect is included, 
one expect the coupling constants to run. It is not clear how to define the infrared limit of the coupling
constant. We can choose an arbitrary  profile for the coupling constant  $e^{2}(z)$ at
the  ultraviolet region and  maybe the effective coupling constant at the low energy may take a universal profile.

We would like to point out some gap between Janus solution in supergravity and Janus field theory.  The maximally  supersymmetric Janus  solution in
supergravity has a limited number of parameters for the dilaton field. This   contrasts to the field theory which can have arbitrary profile of
coupling constants. The coupling constant profile can be regarded as an ultraviolet profile and
the quantum corrections would lead to a change of profile in low energy. However, we do not expect
any universal profile at the low energy as the high energy profile can be chosen to be oscillate. 
Thus we believe that the Janus field theory provides a larger set of theories than those described by the supergravity solution, and would like to find out other alternative origin of Janus field theory.
Also, 
 one could ask which is the exactly  corresponding CFT for the supersymmetric Janus gravity solution. 
 It would be interesting to learn more about both Janus field theory and gravity solution and their relations.

We worked out the cases with the matter fields. One can start from 6-dim theory with
hypermultiplets,  4-dim theory with chiral multiplets, or 3-dim theory with matter multiplets.
The detail will appear soon.

The plan of the paper is as follows. In Sec.2, we review the 8
supersymmetric  Janus Yang-Mills theories. In Sec.3, we study the
vacuum structure of this theory. In Sec.4, we consider the BPS
monopoles and dyons in this theory. In Sec.5, we focus on the
sharp interface for the  the coupling constant. The image charges
for the magnetic monopoles and electric charges are found. The
wave propagation and reflection at the interface is studied. In
Sec.6, less supersymmetric Janus Yang-Mills theories are found
with four real parameters. In Sec.7, we find the supersymmetric
deformation of the Yang-Mills theories when the coupling constant
depends on 2 spacial coordinates. In Sec.8, we find the
supersymmetric deformation in the case where the coupling constant
depends on all three spatial coordinates.

\section{8 Supersymmetric Janus Lagrangian}

The  10-dim supersymmetric Yang-Mills Lagrangian  is
\be {\cal L}_0 = \frac{1}{4e^2} \Tr \Big( -F^{MN}F_{MN}
-2i\bar{\lambda}\Gamma^M D_M \lambda  \Big) ,  \ee
where $M,N=0,1,2,\ldots,9$. We use the 10-dim notation for convenience
with  the gamma matrices $\Gamma^M$ in the Majorana representation
and the gaugino field $\lambda$ is Majorana and Weyl.  The spatial
signature is $(-+++\ldots+)$. The Lagrangian is invariant under the
original supersymmetric transformation
\be \delta_0 A_M=i\bar{\lambda}\Gamma_M\epsilon\, , \, \delta_0\lambda
= \frac{1}{2} \Gamma^{MN}\epsilon F_{MN} , \label{susy0}  \ee
where the Weyl-condition on the susy parameter $\epsilon$ is
\be \Gamma^{012\cdots9}\epsilon=\epsilon .  \label{weyl} \ee
The spinor $\epsilon$ is also a Majorana spinor. As we consider
$1+3$ dim spacetime $x^0,x^1,x^2,x^3$, the remaining  spatial
gradient $\partial_M=0$ with $M=4,5,\ldots,9$ and the gauge field $
A_M $ become scalar fields $\phi_M$ with $M=4,5,\ldots,9$. The
theory has 16 supersymmetries.

In this work, the coupling constant $e^2$ can depend on space-time
coordinates. The original Lagrangian ${\cal L}_0$ transforms as a
total derivative under the original supersymmetric transformation
$\delta_0$ so that
\be \delta_0 {\cal L}_0 = -\partial_\mu\left(\frac{1}{4e^2}\right) 
\Tr \left(
\bar{\lambda}\Gamma^{MN}\Gamma^\mu\epsilon F_{MN}\right) . \label{d0l0}  \ee
Fortunately, one can maintain some of supersymmetries if one
corrects the supersymmetric transformation of the gaugino field by
$\delta_1\lambda$ and also the Lagrangian by additional terms
which depend on the spatial derivatives of the coupling constant.
The additional transformation of the original Lagrangian due to
$\delta_1 \lambda$ would be
\be \delta_1{\cal L}_0 = -\partial_\mu \left(\frac{1}{2e^2}\right)
\Tr \left(
i\bar{\lambda}\Gamma^\mu \delta_1 \lambda -\frac{1}{e^2}
i\bar{\lambda}\Gamma^M D_M\delta_1 \lambda  \label{d1l0}  \right) .
\ee

Let us start with the case where the coupling constant  $e^2$ depends only on the
$x^3=z$ coordinate.  The coupling constant $e^2(z)$ can be an arbitrary function.
 The original 16 supersymmetries should be broken to 8 supersymmetries or
less~\cite{D'Hoker:2006uu}. 
The natural choice of the
additional condition on the spinor $\epsilon$ compatible with the Weyl condition
(\ref{weyl}) is
\be
\Gamma^{3456}\epsilon=\epsilon . \label{susy}\ee
This condition breaks the number of supersymmetries to $8$ and the
global $SO(6)$ symmetry which rotates $4,5,6,7,8,9$ indices to
 $SO(3)\times SO(3)$, each of which
 rotates $4,5,6$ and $7,8,9$ indices respectively.

To cancel some of terms in the zeroth order variation of the
original Lagrangian (\ref{d0l0}), one needs to add a correction to
the susy transformation of the gaugino field and the corrections
to the original Lagrangian. The correction to the original susy
transformation (\ref{susy0}) is
\be \delta_1 A_M=0\; , \;\; \delta_1\lambda = e^2\left(\frac{1}{e^2}\right)'
\sum_{a=4,5,6} \Gamma^{3a}\epsilon  \phi_a ,  \ee
where the prime means $d/dz$. The correction to the original
Lagrangian is made of two parts. The first correction, which
depends on the first order in the derivative of the couple
constant, is given as
\be {\cal L}_1 = \left( \frac{1}{4e^2} \right)' \Tr \Big( i
\bar{\lambda}\Gamma^{456}\lambda - 8i \phi_4[\phi_5,\phi_6] \Big) .
\ee
The second
correction, which is second order in the derivative, is given as
\be {\cal L}_2 =  -
\frac{e^2}{2}\left(\frac{1}{e^2}\right)' \partial_3 \Big(
\frac{1}{e^2} \Tr \sum_{a=4,5,6}  \phi_a^2 \Big) . \ee
The total Lagrangian ${\cal L}= {\cal L}_0+{\cal L}_1+{\cal L }_2$ is
invariant under the corrected susy transformation,
\ba &&  \delta A_M=(\delta_0 +\delta_1) A_M=
i\bar{\lambda}\Gamma_M\epsilon, \nonumber \\
&&  \delta \lambda = (\delta_0 +\delta_1)\lambda =
\frac{1}{2}F_{MN}\Gamma^{MN}\epsilon +e^2\left(\frac{1}{e^2}\right)'
\Gamma^{3a}\epsilon \phi_a .  \label{susy2} \ea
The susy parameter $\epsilon$ is constant in spacetime. There is
no requirement on the space dependence of the coupling constant as
long as it is smooth.

The total Lagrangian ${\cal L}= {\cal L}_0+{\cal L}_1+ {\cal L}_2$
becomes somewhat simpler with change of the field variables as
noted in~\cite{D'Hoker:2006uv}. We divide $\phi_I,\; I=4,5,...,9$
to two groups so that
\be \tilde{\phi}_a \equiv \frac{1}{e^2} \phi_a ,\; (a=4,5,6),
\,\,\, \phi_i= \phi_i \; (i=7,8,9) . \ee
The whole  Lagrangian ${\cal L}$ becomes
\ba {\cal L} &=&\frac{1}{4e^2} \Tr \Big( -F^{\mu\nu}F_{\mu\nu} -
2D^\mu \phi_i D_\mu \phi_i  - 2e^4 D^\mu \tilde{\phi}_a D_\mu
\tilde{\phi}_a \Big) \nonumber \\
& &+ \frac{1}{4e^2}\Tr\Big( [\phi_i,\phi_j]^2
-2e^4[\phi_i,\tilde{\phi}_a]^2+
e^8[\tilde{\phi}_a,\tilde{\phi}_b]^2 \Big) \nonumber
\\ & &
-\frac{i}{2e^2}\Tr \Big( \bar{\lambda}\Gamma^\mu D_\mu \lambda
-i\bar{\lambda}\Gamma^i[\phi_i,\lambda] -ie^2
\bar{\lambda}\Gamma^a[\tilde{\phi}_a,\lambda] \Big) \nonumber \\
& & +  \left(\frac{1}{4e^2}\right)' \Tr \Big(\bar{\lambda}\Gamma^{456}
\lambda - 8i  e^6\tilde{\phi}_4[\tilde{\phi}_5,\tilde{\phi}_6]
\Big). \ea
The combined susy transformation (\ref{susy2}) becomes
\ba && \delta A_\mu = i\bar{\lambda} \Gamma_\mu \epsilon, \;\;
\delta \tilde{\phi}_a = \frac{1}{e^2}\bar{\lambda}\Gamma_a
\epsilon,\;\;
\delta \phi_i = \bar{\lambda}\Gamma_i \epsilon, \nonumber \\
&& \delta \lambda = \Big( \frac{1}{2}F_{\mu\nu}\Gamma^{\mu\nu} + e^2
D_\mu\tilde{\phi}_a\Gamma^{\mu a} + D_\mu\phi_i\Gamma^{\mu i }
\nonumber
\\
&& \;\;\;\;\; -ie^2[\tilde{\phi}_a,\phi_i]\Gamma^{ai}
 -\frac{i}{2}e^4[\tilde{\phi}_a,\tilde{\phi}_b] \Gamma^{ab}
-\frac{i}{2}[\phi_i,\phi_j]\Gamma^{ij}\Big) \epsilon .  \label{susy3} \ea
We can choose the gauge group to be any simple Lie group $G$.

We consider the case where  the coupling constant $e^2(z)$ remain
positive everywhere except  some isolated planes defined by
$z=z_r, r=1,2,...p$ where $e^2(z)$ vanishes. While we expect the
field $\phi_I$ to be  continuous and differentiable everywhere, we
do not expect $\tilde{\phi}_a=\phi_a/e^2$ to be finite and
continuous across the zero planes of the coupling constant. This
would be an important point in the study of the vacuum structure.

If the coupling $e^2(z)$ is an even function of $z$, the
Lagrangian is symmetric under the following $Z_2$ transformation
\be z \rightarrow -z \, , \,\,   A_z\rightarrow -A_z(-z) \, ,
 \,\,  \tilde{\phi}_a\rightarrow -\tilde{\phi}_a \, (a=4,5,6)\, , \,\,
   \lambda\rightarrow \Gamma^{3456} \lambda .
\label{refl}  \ee
On the other hand,  the coupling constant $e^2(z)$ can interpolate
a strong coupling regime with a weak coupling regime. For example,
we can choose the coupling constant profile to be
\be \frac{e^2(z)}{4\pi} = \frac{4\pi}{e^2(-z)} . \ee
The electric coupling and magnetic coupling constants  are exchanged as
one crosses the interface. In this case, the spacial reflection (\ref{refl})  
exchanges the electric and magnetic
sectors.

\section{Vacuum Structure}

Let us consider the minimum of the bosonic energy  density. At the
minimum of the energy, the gauge field strength vanishes and the
gauge field $A_\mu$ is chosen to be zero in a gauge. One can allow
the  usual Coulomb phase  where the scalar fields $\phi_i$ and
$\tilde{\phi}_a$ are homogeneous and diagonal. The Janus theory
may allow additional vacuum structure, as there are corrections to
the original Lagrangian. To see this, let us consider the energy
density for the field $\tilde{\phi}_a$ while showing only $x^3=z$
dependence for the simplicity. The bosonic energy density becomes
\ba {\cal E} &=& \frac{e^2}{2}\Tr \Big( (D_3 \tilde{\phi}_a)^2
-\frac{e^4}{2} [\tilde{\phi}_a,\tilde{\phi}_b]^2 \Big) -i
(e^4)'\Tr\Big(\tilde{\phi}_4[\tilde{\phi}_5,\tilde{\phi}_6] \Big)
\nonumber \\
&=& \frac{e^2}{2} \Tr \Big( D_3 \tilde{\phi}_a +
\frac{e^2}{2}\epsilon_{abc} i[\tilde{\phi}_b,\tilde{\phi}_c]
\Big)^2 -i\Tr \Big(
e^4\tilde{\phi}_4[\tilde{\phi}_5,\tilde{\phi}_6]\Big)' . \ea
Thus the energy functional is bounded below at zero energy if
the boundary term vanishes. The classical vacuum configurations
with zero energy satisfy
\ba && A_\mu=0, \; \partial_\mu\phi_i=0,\;
\partial_{0,1,2}\tilde{\phi}_a=0, \; [\phi_i,\phi_j]=0, \; [\phi_i,\tilde{\phi}_a]=0,
\\
&& D_3\tilde{\phi}_a
+\frac{e^2}{2}\epsilon_{abc}i[\tilde{\phi}_b,\tilde{\phi}_c]= 0 .
\label{vacuumeq}\ea
The last equation is true whenever $e^2\neq 0$.  The vacuum
configurations preserve all the supersymmetries,
 as the gaugino transformation  (\ref{susy3}) becomes
\be \delta \lambda = e^2 \Gamma^{0a}\Big( D_3\tilde{\phi}_a
+\frac{ie^2}{2}\epsilon_{abc}
[\tilde{\phi}_b,\tilde{\phi}_c]\Gamma^{3456} \Big)\epsilon=0 .\ee
The contribution of the boundary term to the energy functional is
given by
\be e^4(z) {\cal F}(z)\Big|^{+\infty}_{- \infty} \  , \ee
where
\be {\cal F}(z)= - i\Tr \big(
\tilde{\phi}_4[\tilde{\phi}_5,\tilde{\phi}_6]\big) . \ee
Using the vacuum equation (\ref{vacuumeq}), we get
\be \frac{d}{dz}{\cal F}(z)  = e^2 \Tr \Big(
-[\tilde{\phi}_4,\tilde{\phi}_5]^2
-[\tilde{\phi}_5,\tilde{\phi}_6]^2
-[\tilde{\phi}_6,\tilde{\phi}_4]^2 \Big)\ge 0,  \ee
and so the  function ${\cal F}(z)$  is non-decreasing  in $z$ in the
interval where $e^2(z)$ is nonvanishing. Thus the boundary term
would not vanish if $e^2(z)$ is nonzero everywhere, and ${\cal F}(z)$ is
nonzero somewhere. However we can have nontrivial nonabelian
vacuum such that  the boundary contributions vanish when  $e^2$
vanishes somewhere, including $z=\pm \infty$.

To solve the vacuum equation (\ref{vacuumeq}), let us introduce a
new variable $u$ such that
\be du =e^2(z) dz, \;\; {\rm or} \;\; u=\int_0^z dz \; e^2(z) .
\label{uint} \ee
In the gauge $A_z=0$, the vacuum equation becomes
\be e^2\Bigg( \frac{d\tilde{\phi}_a}{du}+
 \frac{i}{2}\epsilon_{abc}[\tilde{\phi}_b,\tilde{\phi}_c] \Bigg)= 0 . \ee
When $e^2\neq 0$, the above equation is the Nahm equation for
magnetic monopoles\cite{Nahm:1979yw}. However at points where
$e^2(z)=0$, the Nahm equations does not need to hold. As before we
assume that $e^2(z)$ vanishes at finite number of points $z_r$,
and we divides the $z=x^3$ line into finite number of intervals
separated by zero points $z_r$.  The fields $\tilde{\phi}_a$ need
not be continuous nor finite at these zero points as long as the
original unscaled field $\phi_a$ is so. Thus we are solving the
Nahm at each interval. For each interval between zero coupling
constant points $z_r$, one has to impose the Nahm equations in $u$
variables. In addition we require the  contribution of the boundary term
to be finite,  continuous at $z_r$, and vanishes at $\pm \infty$.

To be more concrete let us focus on the gauge group $SU(2)$. The
general solutions of the Nahm equation can be obtained by using
the ansatz,
\be \tilde{\phi}_{3+a} = f_a(u) \frac{\sigma_a}{2}  \ee
with the Pauli matrices $\sigma_a$ and no sum over the indices
$a=1,2,3$. The vacuum equation becomes
\be f'_1=f_2 f_3,\;\; f'_2= f_3 f_1,
\;\; f'_3= f_1f_2 ,  \ee
whose solutions are given in terms of the Jacobi elliptic functions, as follows:
\ba && f_1(u;k,D,u_0) \equiv -\frac{D{\rm cn}_k[D(u-u_0)]}{{\rm sn}_k[D(u-u_0)]} ,
 \nonumber \\
&&  f_2(u;k,D,u_0)  \equiv -\frac{D{\rm dn}_k[D(u-u_0)]}{{\rm sn}_k[D(u-u_0)]} ,
 \nonumber \\
&& f_3(u;k,D,u_0)  \equiv - \frac{D}{{\rm sn}_k[D(u-u_0)]} ,
\label{nahm} \ea
where $k\in [0,1]$ is the elliptic modulus, and two parameters
$D\ge 0,\;  u_0$ are arbitrary. This solution blows up when ${\rm
sn}_k$ goes to zero. The zeros of ${\rm sn}_k(w)$ is $w=0, 2K(k)$
where $K(k)$ is the complete elliptic integral of the first kind.
The function  $K(k)$ goes to infinite at the boundary $k=1$. The
above solution in this limit becomes
\ba && f_1(u;k=1,D,u_0) =
-\frac{D\cosh(D(u-u_0))}{\sinh(D(u-u_0))}, \nonumber \\
&& f_2(u;k=1,D,u_0) = f_3(u;k=1,D,u_0) =-\frac{D}{\sinh D(u-u_0)} .
\label{semi}\ea
When $D\neq 0$ nor $K(k)=\infty$, the general solution
(\ref{nahm}) blows up at finite $u$. If there is no  point
including infinities where $e^2$ vanishes, one can see there is no
nontrivial vacuum solution.

Let us now consider the case where $e^2$  vanishes only one point, say
at $z=0$, and remain positive and finite everywhere else. We do not
need the detail profile of 
the coupling constant $e^2(z)$ for our discussion. The parameter
$u$ in Eq.(\ref{uint}) is negative for $z<0$ and positive for
$z>0$. We have two semi-infinite intervals and so need the
above solution (\ref{semi}) for these two intervals. We could
choose independent parameters for two interval and so the vacuum
solution becomes    
\be  \tilde{\phi}_{3+a}=\left\{ \begin{array}{ccc}
f_a(u;k=1,D_-,u_-)\frac{\sigma_a}{2} & {\rm for} & z<0 \\
f_a(u;k=1,D_+,u_+)\frac{\sigma_a}{2} & {\rm for} & z>0
\end{array} , \label{vacuum0} \right.
\ee
where $u_->0, u_+<  0$. The range of two parameters $u_\pm$ is
chosen so that $\tilde{\phi}_a$ does not diverge anywhere. If we
have chosen $u_-=0$, we would have divergent contribution to the
boundary term at $0_-$ as $\phi_a \sim 1/(e^2(t)t)\sigma_a$ near
$z=0_-$. The asymptotic values of $\tilde{\phi}$ at the spatial
infinity becomes
\be \tilde{\phi}_a(z=\pm \infty) = - \delta_{a4} D_\pm
\frac{\sigma_1}{2} . \ee
Not only the asymptotic value $D_\pm$ can be different, they can
vanish. Thus, one can have nontrivial vacuum even in the symmetric
phase. The above solution (\ref{vacuum0}) becomes abelian in
asymptotic region $(z=\pm \infty)$ but nonabelian close to the
zero plane $z=0$. The $SU(2)$  gauge symmetry is completely broken
near the wall but becomes abelian when $D_\pm\neq 0$ or fully
restored  when $D_\pm=0$ at the boundaries $z=\pm\infty$.

When there are more planes where $e^2(z)$ vanishes, one can have
a richer vacuum structure. For each finite interval between zeros,
the full general solution (\ref{nahm}) will play a role. The above
solution (\ref{vacuum0}) becomes abelian in asymptotic region $(z=\pm \infty)$ but
nonabelian close to the zero planes $z=z_r$. The $SU(2)$  gauge
symmetry is completely broken near the wall but becomes abelian or
fully restored at the boundaries. There are several parameters
characterizing the vacuum, besides the global $SU(2)$ rotation of
three scalar fields $\tilde{\phi}_a$. The detailed physics in a
given vacuum is intriguing but will not be pursued in this work.

\section{BPS Objects}

The BPS configurations are those which respect some
supersymmetries. Let us consider the supersymmetric transformation
(\ref{susy2})  of the gaugino field. In each vacuum one can study
the BPS configurations. The supersymmetry preserved by the BPS
configurations should be compatible with the original
supersymmetric condition, $\Gamma^{3456}\epsilon=\epsilon$. We
will consider the following two  conditions on the supersymmetric
parameter, $\epsilon$;
\be \Gamma^{1234}\epsilon=\alpha \epsilon,\;\; \Gamma^{07}\epsilon = \beta
\epsilon ,  \label{bpss}\ee
where $\alpha=\pm 1, \beta=\pm 1$.   The above relations imply that
$\Gamma^{1256}\epsilon=-\alpha \epsilon$,
$\Gamma^{1289}\epsilon=\beta \epsilon$, and
$\Gamma^{5689}\epsilon=\alpha\beta\epsilon$. We could impose only one
condition and then the configurations would be 1/2 BPS. If we
impose both conditions, the configurations would be 1/4 BPS. 

One may wonder whether there are other possible BPS conditions.
 As the fields are Majorana, we cannot introduce, for example, the projection $\Gamma^{12}\epsilon=i\epsilon$. 
Other possible projections like  $\Gamma^{1256}\epsilon=\epsilon$ or $\Gamma^{1289}\epsilon=\epsilon$
are allowed.  But these conditions would lead to the reduction of the selfdual Yang-Mills
equation to 2-spatial direction, which does not have any obvious nontrivial smooth solution. 
The above BPS conditions (\ref{bpss}) are those for  magnetic monopoles and charged W-bosons 
in non-Janus case and might imply  nontrivial BPS configurations even in the Janus case. 

The supersymmetric transformation (\ref{susy2}) of the gaugino field
can be expressed as
\ba && \!\!\!\!\!\!\!\! \delta \lambda =
\Gamma^{p0}(F_{p0}-D_p\phi_7 \Gamma^{07}) \epsilon +
e^2\Gamma^{0a}(D_0\tilde{\phi}_a +i[\phi_7,\tilde{\phi}_a ]
\Gamma^{07})\epsilon
+\sum_{i=8,9}\Gamma^{0i}(D_0\phi_i+i[\phi_7,\phi_i]\Gamma^{07})\epsilon
\nonumber \\
&& +\Gamma^{12}(F_{12}-e^2 D_3\tilde{\phi}_4 \Gamma^{1234} +e^4
i[\tilde{\phi}_5,\tilde{\phi}_6]\Gamma^{1256}
+i[\phi_8,\phi_9]\Gamma^{1289})\epsilon +
\Gamma^{23}(F_{23}-e^2D_1\tilde{\phi}_4\Gamma^{1234}) \epsilon
\nonumber \\
&& + \Gamma^{31}(F_{31}-e^2D_2\tilde{\phi}_4\Gamma^{1234})\epsilon
+e^2 \Gamma^{15}(D_1
\tilde{\phi}_5+D_2\tilde{\phi}_6\Gamma^{1256})\epsilon
+e^2\Gamma^{25}(D_2\tilde{\phi}_5-D_1\tilde{\phi}_6\Gamma^{1256})\epsilon
\nonumber \\
&&
+e^2\Gamma^{35}(D_3\tilde{\phi}_5+ie^2[\tilde{\phi}_6,\tilde{\phi}_4]\Gamma^{3456}
)\epsilon + e^2\Gamma^{36}
(D_3\tilde{\phi}_6+ie^2[\tilde{\phi}_4,\tilde{\phi}_5]
\Gamma^{3456})\epsilon \nonumber \\
&&+  \Gamma^{18}(D_1\phi_8+D_2\phi_9\Gamma^{1289})\epsilon
+\Gamma^{28}(D_2\phi_8-D_1\phi_9\Gamma^{1289})\epsilon \nonumber \\
&&+\Gamma^{38}(D_3\phi_8-ie^2[\tilde\phi_4,\phi_9]\Gamma^{3489})\epsilon
+\Gamma^{39}(D_3\phi_9+ie^2[\tilde\phi_4,\phi_8]\Gamma^{3489})\epsilon
\nonumber \\
&&+ e^2\Gamma^{58}(-i[\tilde{\phi}_5,\phi_8]
-i[\tilde{\phi}_6,\phi_9]\Gamma^{5689})\epsilon
+ e^2\Gamma^{59}(-i[\tilde{\phi}_5,\phi_9]+i[\tilde{\phi}_6,\phi_8]
\Gamma^{5689})\epsilon \nonumber \\
&& + D_0\phi_7 \Gamma^{07} \epsilon. \ea
The susy transformation $\delta \lambda$ would vanish for the BPS
configurations. After using the  BPS conditions (\ref{bpss}), $\delta \lambda=0$ if all terms vanish
individually.  (It would be interesting to show that it is also a necessary condition.)
  Let us consider the magnetic 1/2 BPS equation with $\alpha=1$.  We require all terms
  vanish with $\beta=\pm 1$.   The nontrivial part of the  equations for the 1/2 BPS configurations
with $\Gamma^{1234}\epsilon = \epsilon$ is made of 
\ba && F_{12} -e^2 D_3\tilde{\phi}_4 -
ie^4[\tilde{\phi}_5,\tilde{\phi}_6]= 0, \;\;  F_{23} -e^2
D_1\tilde{\phi}_4=0,\;\; F_{31}-e^2
D_2\tilde{\phi}_4=0, \nonumber \\
&& D_{3}(\tilde{\phi}_5 + i \tilde{\phi}_6) - e^2
[\tilde{\phi}_4,\tilde{\phi}_5+i\tilde{\phi}_6]=0,\;\; (D_1+iD_2)
(\tilde{\phi}_5+i\tilde{\phi}_6)=0.
\label{bps01}  \ea
This  is a mixed form of the Nahm equation for the vacuum and the old BPS equation
for  magnetic monopoles.
The 1/4 BPS  dyonic
magnetic monopole with $\beta=1$ can found also. 
 The additional BPS equation for dyons in
the  gauge $A_0=\phi_7$ and the ansatz $\phi_8=\phi_9=0$ is simply
the Gauss law,
\be -D_p\Big(\frac{1}{e^2}D_p \phi_7\Big)+e^2
[\tilde{\phi}_a,[\tilde{\phi}_a,\phi_7]] = 0 . \label{bps02}
 \ee
In the abelian Coulomb phase, $\tilde{\phi}_5=\tilde{\phi}_6=0$
and the above BPS equations become somewhat simpler.
(Of course it would be interesting to find whether there is nontrivial BPS configurations
lying beyond the ansatz $\phi_=\phi_9=0$. )

For simplicity, let us consider the energy bound in the abelian
Coulomb vacuum. Keeping only nontrivial terms, we express  the
energy functional as
\be {\cal H} = \int d^3x \; \frac{1}{2e^2} \Tr \Big( (F_{p0}- D_p
\phi_7)^2 + e^4(D_0\tilde{\phi}_4- i[\phi_7,\tilde{\phi}_4])^2+
(B_p -e^2D_p\tilde{\phi}_4)^2 \Big) + Q_e + Q_m, \ee
where $B_p = \frac12\epsilon_{pkl}F^{kl}$ and
\be Q_e= \int d^3x \partial_p \Tr
\Big(\frac{1}{e^2}F_{p0}\phi_7\Big) ,\;\;\;\; Q_m=\int d^3x
\partial_p \Tr (B_p\tilde{\phi}_4) ,  \ee
are the electric and magnetic energy contributions, respectively.
In the Janus field theory, the coupling constant $e^2$ depending
on the spatial coordinates and so it is much harder to solve the
BPS equations even for a single magnetic monopole. Magnetic
monopoles are topologically characterized in usual abelian vacuum,
but it is not clear whether it is so in a nonabelian vacuum.

\section{A Sharp Interface}

\subsection{BPS monopoles and point electric charge}
			
Suppose the
coupling constant $e^2(z) $ changes from one value to another at a
sharp interface so that 
\be e(z) = \left\{ \begin{array}{cc}
 e_1 & {\rm  for }\;\; z>0 \\
 e_2 & {\rm for} \;\; z<0 \end{array}\right.\  . \label{interface} \ee
Such a limit can be obtained by shrinking the
interface region to a plane.  As there is 
no additional source term at the interface,
we get the continuity conditions of the various fields. The continuous 
ones are the following fields and their covariant derivatives:  
\ba &&  F_{01}, F_{02}, F_{12}, \frac{F_{03}}{e^2},
\frac{F_{23}}{e^2}, \frac{F_{31}}{e^2}, \nonumber \\
&& \tilde{\phi}_a, D_1 \tilde{\phi}_a, D_2\tilde{\phi}_a, e^2
D_3\tilde{\phi}_a ,\quad a=4,5,6   \nonumber   \\
&&  \phi_i, D_1 \phi_i, D_2 \phi_i, \frac{D_3\phi_i}{e^2},\quad
i=7,8,9.   \label{bdcondition} \ea
Thus naturally we can assume the continuity condition for
the  infinitesimal gauge function $\Lambda$ and its derivatives
$D_1 \Lambda, D_2 \Lambda, D_3\Lambda/e^2$.

For simplicity, we consider the $SU(2)$ gauge theory which is
broken spontaneous to $U(1)$ subgroup by the Higgs expectation
values at the vacuum,
\be <\tilde{\phi}_4>=\tilde{v}\frac{\sigma_3}{\sqrt{2}} .  \ee
Note that the expectation value of the  original field variable
$\phi_4=e^2\tilde{\phi}_4$ makes a jump at the interface. The
diagonal components of the fields will be massless and
off-diagonal fields will be massive. Let us try to solve the BPS
equations in the abelian limit where the nonabelian core size
vanishes.  For a single monopole
at $z=a>0$, we get the BPS configuration 
\be  B_i = e^2 D_i\tilde{\phi}_4 =\left\{ \begin{array}{cc} 
 \frac{(x,y,z-a)}{r^3_+} +
\frac{e_1^2-e_2^2}{e_1^2+e_2^2} \frac{(x,y,z+a)}{r_-^3} \;\;
& , \; z>0  \\
  \frac{2e_2^2}{e_1^2+e_2^2} \frac{(x,y,z-a)}{r_+^3} \;\;
& ,\;  z<0  \end{array}\right. \ . \ee
Here we dropped the group factor $\sigma_3/\sqrt{2}$ for the simplicity. 
The continuous scalar field $\tilde{\phi}_4$ becomes
\be \tilde{\phi}_4= \left\{ \begin{array}{cc}
 \tilde{v} -\frac{1}{e_1^2 r_+} -
\frac{e_1^2-e_2^2}{e_1^2(e_1^2+e_2^2)} \frac{1}{r_-} &,\;  z>0 \\
 &  \\
\tilde{v}-\frac{2}{e_1^2+e_2^2} \frac{1}{r_+} &, \;  z<0
\end{array} \right. \ . \ee
The total magnetic flux near $z=a$ is $4\pi$ as expected. In the
region $z>0$ where the monopole exists, the total field is that of
the magnetic monopole and that of the mirror image at $z=-a$. The
total magnetic flux $4\pi$ at the spacial infinity consists of the
$4 \pi e_1^2/(e_1^2+e_2^2)$ flux from the $z>0$ hemisphere and the
$4\pi e_2^2/(e_a^2+e_2^2)$ flux from the $z<0$ hemisphere.

Let us now turn off the $\tilde{\phi}_4$ expectation value and
turn on the new expectation value
\be  <\phi_7>= u\frac{\sigma_3}{\sqrt{2}} . \ee
Let us put an unit electric charge at point $(x,y,z) = (0,0,a>0)$.
The Gauss law is simplified as $\nabla_i (E_i/e^2)=\rho_e$ whose
spatial integration is quantized as integer. Ignoring the
nonabelian core and dropping the group factor $\sigma_{3}/\sqrt{2}$ for the
simplicity,  we get the BPS point charge configuration as
\be E_i=D_i \phi_7 = \left\{ \begin{array}{cc}   \frac{e_1^2}{4\pi} \;\left(
\frac{(x,y,z-a)}{r_+^3} +
\frac{-e_1^2+e_2^2}{e_1^2+e_2^2}\; \frac{ (x,y,z+a)}{r_-^3} \right) &  , \,  z>0 
 \\
 \frac{e_1^2}{4\pi}\; \frac{2 e_2^2}{e_1^2+e_2^2} \; \frac{
(x,y,z-a)}{r_+^3} & ,\,  z<0 \end{array}\right. \ .\ee 
where $r_\pm^2 = x^2+y^2+(z\mp a)^2 $. Note that $E_1, E_2,
E_3/e^2$ are continuous along the interface. The continuous scalar
field becomes
\be \phi_7= \left\{ \begin{array}{cc}
 u -\frac{e_1^2}{4\pi}\left( \frac{1}{r_+}+\frac{-e_a^2+e_2^2}{e_1^2+e_2^2}
 \frac{1}{r_-}\right) &, \,  z>0 \\
 &  \\
u-\frac{2e_1^2e_2^2}{4\pi(e_1^2+e_2^2)} \frac{1}{r_+} &, \,  z<0
\end{array} \right. \ .  \ee
The total electric charge is the unity near $z=1$ and remains so at the
spatial infinity as it is  the sum $
e_2^2/(e_1^2+e_2^2) , (z>0)$ and $e_1^2/(e_1^2+e_2^2), z<0$.

\subsection{Reflection and  transmission of massless waves} 

Let us consider now a massless wave propagating toward the interface
(\ref{interface}) of the two coupling constant from $z>0$ region. 
The fields and their derivatives in (\ref{bdcondition}) should be continuous
cross the interface $z=0$. Let us use the vector notation ${\bf E}= (F_{10},
F_{20}, F_{30})$, and ${\bf B}=(F_{23}, F_{31}, F_{12})$ for the electromagnetic
fields.   A part of the incident wave will be reflected and the 
rest may get refracted or transmitted. Let us
call the electromagnetic field of the incident wave to be ${\bf
E},{\bf B}$, the reflected wave to be ${\bf E}'', {\bf B}''$ and
the transmitted wave to be ${\bf E}',{\bf B}'$. The continuity equations at $z=0$ are
\ba && \big({\bf E}+ {\bf E}''-{\bf E}'\big) \times \hat{z}= 0 ,
\nonumber
\\
&& \big( {\bf B}+{\bf B}''-{\bf B} \big) \cdot \hat{z} = 0 ,
\nonumber \\
&& \bigg( \frac{ {\bf E}+{\bf E}''}{e_1^2} -\frac{{\bf
E}'}{e_2^2} \bigg) \cdot \hat{z}= 0 ,\nonumber \\
&& \bigg(\frac{{\bf B}+{\bf B}''}{e_1^2} -\frac{{\bf B}'}{e^2_2}
\bigg) \times \hat{z} = 0. \label{conti2} \ea
The space-time dependence waves would be  $e^{-iwt+{\bf k}\cdot
{\bf x}}$, $e^{-iwt+{\bf k}''\cdot {\bf x}}$, and $e^{-iwt+{\bf
k}'\cdot {\bf x}}$ for the incident, reflected, and transmitted
waves, respectively. The wave equation at each region and the
above continuity equations imply that
\be w= |{\bf k}|=|{\bf k}''|=|{\bf k}'|, \;\; {\bf k}={\bf k}' ,
\;\;  ({\bf k}+{\bf k}')\times \hat{z}=0 . \ee
Thus the transmitted wave is not refracted at all. After taking
out the space-time dependence,  we can express the electric fields
of the reflected and transmitted waves in terms of the electric
field of the incident wave. While the relation will depends on
whether the wave has transverse electric (that is, transverse to
the incident plane defined by ${\bf k}$ and $\hat{z}$), or
transverse magnetic, both cases has the same relation
between the magnitude of the electric field at $z=0$, as $E_0'' =
r E_0, E_0'= t E_0$ where the reflection and transmission
magnitudes are
\be r = \left|\frac{e_1^2-e_2^2}{e_1^2+e_2^2}\right|,\;\;
t=\frac{2e_2^2}{e_1^2+e_2^2} . \ee
For the vector, one should be careful about the sign, which 
can be easily fixed  by the continuity equations. The same reflection and
transmission magnitudes apply to the scalar fields
$\phi_i,i=7,8,9$. For the scalar field $\tilde{\phi}_a$, the same
reflection magnitude applies but the transmission magnitude
becomes $t=2e_1^2/(e_1^2+e_2^2)$.

\section{Additional Susy Breaking Janus}

In this section we are still interested in the
case where the coupling constant $e^2(z)$ depends only on one
spatial coordinate. We can impose additional constraints on the
susy parameters $\epsilon$ which is compatible with what we have
already imposed. There are several of them and so one can break
the susy to 1/4 or 1/8, which introduces some free parameters in
the interface Lagrangian. We easily recover the results in Ref.~\cite{D'Hoker:2006uv}.
As shown in this reference, our study exhaust all possibilities
with some supersymmetries. 
Thus the minimum one will have two supersymmetries for the
case where the coupling constant depends only on one spatial
direction $e^2(z)$.
The compatible conditions including one in (\ref{susy}) on
the 10-dim Majorana Weyl spinor $\epsilon$
are
\be \Gamma^{3456}\epsilon=\epsilon\, , \; \Gamma^{3489}\epsilon=
-\epsilon\, , \;   \Gamma^{3597}\epsilon=-\epsilon\, , \;
\Gamma^{3678}\epsilon=-\epsilon . \label{1dim8}\ee
As the product of the above four conditions is an identity, there
are only three independent conditions, breaking the supersymmetry
to 1/8th or two supersymmetries.

To cancel $\delta_0{\cal L}_{0}$ in (\ref{d1l0}), we choose  the first
correction to the Lagrangian to be
\ba  {\cal L}_1 \!\!\! &=& \!\! \left(\frac{1}{4e^2}\right)' \Tr
\Big( i \bar{\lambda} (c_0\Gamma^{456} - c_1\Gamma^{489}-
c_2\Gamma^{597}-
c_3\Gamma^{678} )\lambda \nonumber \\
& &  \!\!\!\! \!\!\!\!\! -8i \Big(c_0\phi_4[\phi_5,\phi_6]
-c_1\phi_4[\phi_8,\phi_9] - c_2\phi_5[\phi_9,\phi_7] -c_3
\phi_6[\phi_7,\phi_8])\Big), \ea
where real parameters $c_i$ satisfy
\be c_0+ c_1+c_2+c_3=1. \ee
The correction to the susy transformation (\ref{susy0}) is
\ba \delta_1\lambda &=& e^2\left(\frac{1}{e^2}\right)' \Gamma^3
\Big( c_0\sum_{a=4,5,6}\Gamma^a\phi_a +
c_1\sum_{a=4,8,9} \Gamma^a\phi_a \nonumber \\
&& \;\;\;\; + c_2\sum_{a=5,9,7}\Gamma^a\phi_a
 +
c_3\sum_{a=6,7,8}\Gamma^a\phi_a \Big)\epsilon. \ea
The second order correction of the Lagrangian is chosen to be 
\ba && {\cal L}_2=
-\frac{e^2}{2}\left(\frac{1}{e^2}\right)'\partial_3\Big(
\frac{1}{e^2} \Tr \Big( c_0\sum_{a=4,5,6}\phi_a^2 +
c_1\sum_{a=4,8,9}\phi_a^2 +c_2\sum_{a=5,9,7}\phi_a^2 \nonumber \\
&& \;\;\;\;  +c_3 \sum_{a=6,7,8}\phi_a^2 \Big)\Big)
+\frac{e^2}{2}\left(\frac{1}{e^2}\right)'^2 \Tr \Big(
(c_0+c_1)(c_2+c_3)(\phi_4^2+\phi_7^2) \nonumber \\
&& \;\;\;\; + (c_0+c_2)(c_1+c_3)(\phi_5^2+\phi_8^2) +
(c_0+c_3)(c_1+c_2)(\phi_6^2+\phi_9^2) \Big) . \ea

As noted in Ref.~\cite{D'Hoker:2006uv}, notice
that when $c_0=c_1=c_2=c_3=1/4$, there is an enhanced global
symmetry $SU(3)$ with 1/8 supersymmetry.  For $c_0=c_1=1/2$ and
$c_2=c_3=0$, there is 1/4 supersymmetry with enhanced global
symmetry $SO(2)\times SO(2)$.

\section{Multifaced Interfaces in 2,3 Dimensions}

\subsection{$e^{2}(y,z) $ case}

Let us first start with the case where the
coupling constant $e^2(y,z)$ depends on only two coordinates.
There exist only two independent, modulo rotation, 
sets of the compatible supersymmetry
conditions which are
\ba && \,  \Gamma^{2789}\epsilon=\epsilon  , \;
\Gamma^{3456}\epsilon=\epsilon   ,\;\;\; \label{inter1}\\
&&  \, \Gamma^{2459}\epsilon=-\epsilon , \;
\Gamma^{3456}\epsilon=\epsilon   .  \label{inter2}  \ea
Each condition breaks the supersymmetry to 1/4.  One can break the
supersymmetry further to 1/8 by imposing both conditions
(\ref{inter1}) and (\ref{inter2}) at the same time. Also one can
impose additional compatible supersymmetry condition
\be  \,  \Gamma^{2567}\epsilon=-\epsilon  , \
\Gamma^{3456}\epsilon=\epsilon .  \label{inter3}  \ee
Imposing these three mutually independent and compatible
conditions (\ref{inter1}), (\ref{inter2}), (\ref{inter3}) breaks
 the supersymmetry
to the minimal one 1/16.  Note that the conditions (\ref{inter2}) and (\ref{inter3}) 
are  related by a rotation.
These three conditions imply
\be
\Gamma^{2648}\epsilon=\Gamma^{3489}\epsilon
=\Gamma^{3597}\epsilon=\Gamma^{3678}\epsilon=-\epsilon .
\ee
These conditions  include the conditions (\ref{1dim8}) in the
previous section.

We extend the result in the previous section. To cancel $\delta_0 {\cal L}_{0}$, we choose
choose the first order correction to the Lagrangian to be
\ba  {\cal L}_1 \!\!\! &=& \!\!
\partial_2\left(\frac{1}{4e^2}\right) \Tr \Big( i \bar{\lambda}
(b_0\Gamma^{789} - b_1\Gamma^{567}- b_2\Gamma^{648}-
b_3\Gamma^{459} )\lambda \nonumber \\
& &  \!\!\!\! \!\!\!\!\! -8i \Big(b_0\phi_7[\phi_8,\phi_9]
-b_1\phi_5[\phi_6,\phi_7] - b_2\phi_6[\phi_4,\phi_8] -b_3
\phi_4[\phi_5,\phi_9])\Big) \nonumber \\
 &+& \!\!
\partial_3\left(\frac{1}{4e^2}\right) \Tr \Big( i \bar{\lambda}
(c_0\Gamma^{456} - c_1\Gamma^{489}- c_2\Gamma^{597}-
c_3\Gamma^{678} )\lambda \nonumber \\
& &  \!\!\!\! \!\!\!\!\! -8i \Big(c_0\phi_4[\phi_5,\phi_6]
-c_1\phi_4[\phi_8,\phi_9] - c_2\phi_5[\phi_9,\phi_7] -c_3
\phi_6[\phi_7,\phi_8])\Big), \ea
where real parameters $b_i, c_i$ satisfy
\be b_0+b_1+b_2+b_3=1  , \,\, c_0+ c_1+c_2+c_3=1  . \ee
The correction to  the supersymmetric transformation (\ref{susy}) is
\ba && \delta_1\lambda = e^2\partial_2 \left(\frac{1}{e^2}\right)
\Gamma^2 \Big( b_0\! \sum_{a=7,8,9}\Gamma^a\phi_a + b_1\!
\sum_{a=5,6,7}\Gamma^a\phi_a + b_2\! \sum_{a=6,4,8}\Gamma^a\phi_a
 +
b_3\! \sum_{a=4,5,9} \Gamma^a\phi_a \Big)\epsilon
\nonumber \\
&& \!\!\!\!\!  \!\!\!\! +e^2\partial_3 \left(\frac{1}{e^2}\right)
\Gamma^3 \Big( c_0\! \sum_{a=4,5,6}\Gamma^a\phi_a + c_1\!
\sum_{a=4,8,9}\Gamma^a\phi_a + c_2\! \sum_{a=5,9,7}\Gamma^a\phi_a
+ c_3\! \sum_{a=6,7,8}\Gamma^a\phi_a \Big)\epsilon.  \ea
The additional correction to the Lagrangian is made of
\ba && {\cal L}_2= -\frac{e^2}{2}\partial_2
\left(\frac{1}{e^2}\right) \partial_2 \Bigg( \frac{1}{e^2} \Tr
\Big( b_0\sum_{a=7,8,9}\phi_a^2 + b_1\sum_{a=5,6,7}\phi_a^2 +
b_2\sum_{a=6,4,8}\phi_a^2 +b_3\sum_{a=4,5,9}\phi_a^2 \Big)\Bigg)
\nonumber \\
&& \!\!\!\!\!\!  -\frac{e^2}{2}\partial_3 \left(\frac{1}{e^2}\right)
\partial_3 \Bigg( \frac{1}{e^2} \Tr \Big(
c_0\sum_{a=4,5,6}\phi_a^2 + c_1\sum_{a=4,8,9}\phi_a^2 +
c_2\sum_{a=5,9,7}\phi_a^2 +c_3\sum_{a=6,7,8}\phi_a^2 \Big)\Bigg).
\ea
One needs additional correction to the Lagrangian which are made
of mixed terms,
\ba {\cal L}_3 &=&
\frac{e^2}{2}\left(\partial_2\Big(\frac{1}{e^2}\Big)\right)^2 \Tr \Big(
(b_0+b_1)(b_2+b_3)(\phi_7^2+\phi_4^2) +
 (b_0+b_2)(b_1+b_3)(\phi_8^2+\phi_5^2) \nonumber \\
& &   +
(b_0+b_3)(b_1+b_2)(\phi_9^2+\phi_6^2) \Big)
+\frac{e^2}{2}\left(\partial_3\Big(\frac{1}{e^2}\Big)\right)^2 \Tr \Big(
(c_0+c_1)(c_2+c_3)(\phi_4^2+\phi_7^2) \nonumber \\
& &
 + (c_0+c_2)(c_1+c_3)(\phi_5^2+\phi_8^2) +
(c_0+c_3)(c_1+c_2)(\phi_6^2+\phi_9^2) \Big) \nonumber \\
& & -\left( \partial_2\partial_3\Big(\frac{1}{e^2}\Big) -e^2\partial_2
\Big(\frac{1}{e^2}\Big)\partial_3\Big(\frac{1}{e^2} \Big)\right)
\Tr \Big( (b_0+b_1+c_0+c_1-1)\phi_4\phi_7\nonumber\\
& & +(b_0+b_2+c_0+c_2-1)\phi_6\phi_8 + (b_0+b_3+c_0+c_3-1)\phi_7\phi_9  \Big) .  \ea
The total Lagrangian ${\cal L}_0 +{\cal L}_1 +{\cal L}_2 +{\cal L}_3$ is invariant
under the corrected supersymmetric transformation.
Note that when $e^2=f(y)g(z)$ so that it is factorizable, the last term vanishes.
When $b_a=c_a= 1/4$ for all $a=0,1,2,3$, there is $SO(3)$ 
symmetry which rotates $\phi_{4,5,6}$ and $\phi_{7,8,9}$ 
at the same time. We think our Lagrangian is the most general on in the 2-dim case.

\subsection{$e^{2}(x,y,z)$ case}

When the coupling constant depends on all three coordinates
$e^2(x,y,z)$, there are two independent susy conditions
\ba &&  \Gamma^{1467}\epsilon=\Gamma^{2475}\epsilon=
\Gamma^{3456}\epsilon=\epsilon,  \label{3dima}
\\
&& \Gamma^{1458}\epsilon=\Gamma^{2468}\epsilon=
\Gamma^{3478}\epsilon=\epsilon . \label{3dimb}  \ea
One can break either susy further to 1/16 by imposing the above
two conditions (\ref{3dima}) and (\ref{3dimb}) together. These two
conditions  imply
$\Gamma^{1234}\epsilon=\Gamma^{5678}\epsilon=-\epsilon$.
We choose the first correction to the Lagrangian to be
\ba && \!\! \!\! {\cal L}_1 = \partial_1 \left(\frac{1}{4e^2}\right) \Tr \Big(
i\bar{\lambda}\Gamma^4(a_1 \Gamma^{67}+a_2\Gamma^{58})
 \lambda -8i\phi_4(a_1 [\phi_6,\phi_7]
+a_2[\phi_5,\phi_8])\Big) \nonumber \\
&&  \!\! \!\! +
\partial_2 \left(\frac{1}{4e^2}\right)
\Tr \Big(i\bar{\lambda}\Gamma^4(b_1\Gamma^{75}+b_2\Gamma^{68})
\lambda- 8i\phi_4(b_1 [\phi_7,\phi_5]) +b_2[\phi_6,\phi_8]) \Big)
\nonumber  \\  &&  \!\! \!\! +
\partial_3 \left(\frac{1}{4e^2}\right) \Tr \Big((i\bar{\lambda}\Gamma^4(c_1
\Gamma^{56}+c_2\Gamma^{78})
\lambda -8i\phi_4(c_1 [\phi_5,\phi_6]+c_2[\phi_7,\phi_8]) \Big), \ea
where
\be a_1 + a_2=1 , \, b_1+b_2=1   , \, c_1+c_2=1. \ee
We choose the correction for the susy transformation to be
\ba   \delta_1 \lambda &=&
e^2\partial_1\left(\frac{1}{e^2}\right)\Gamma^1 \Big( a_1 \sum_{a=4,6,7}
\phi_a\Gamma^{a}+ a_2\sum_{4,5,8}\phi_a \Gamma^{a} \Big)\epsilon \nonumber \\
&+&
e^2\partial_2\left(\frac{1}{e^2}\right)\Gamma^2 \Big( b_1 \sum_{a=4,7,5}
\phi_a \Gamma^{a}+ b_2 \sum_{a=4,6,8}\phi_a\Gamma^a \Big)\epsilon \nonumber \\
&+&
e^2\partial_3\left(\frac{1}{e^2}\right)\Gamma^3 \Big( c_1\sum_{a=4,5,6}\phi_a
\Gamma^{a}+ c_2\sum_{a=4,7,8}\phi_a\Gamma^a\Big) \epsilon . \ea

The additional Lagrangian becomes
\ba && {\cal L}_2= -\frac{e^2}{2} \partial_1
\left(\frac{1}{e^2}\right)\partial_1 \Tr
\big( a_1 (\phi_4^2+\phi_6^2+\phi_7^2) +a_2 (\phi_4^2+\phi_5^2+\phi_8^2)\big)
 \nonumber \\
&&   -\frac{e^2}{2}\partial_2
\left(\frac{1}{e^2}\right)\partial_2 \Tr  \big( b_1
(\phi_4^2+\phi_5^2+\phi_7^2) +b_2 (\phi_4^2+\phi_6^2+\phi_8^2)\big)  \nonumber \\
&& -\frac{e^2}{2}\partial_3
\left(\frac{1}{e^2}\right)\partial_3 \Tr \big(c_1
(\phi_4^2+\phi_5^2+\phi_6^2) +  c_1 (\phi_4^2+\phi_7^2+\phi_8^2)\big). \ea
The final mixed correction to the Lagrangian is
\ba {\cal L}_3= && \frac{e^2}{2}\Bigg( a_1a_1 \left(\partial_1
\Big(\frac{1}{e^2}\Big) \right)^2     +b_1 b_2
\left(\partial_2 \Big(\frac{1}{e^2}\Big) \right)^2 +c_1c_2
 \left(\partial_3 \Big(\frac{1}{e^2}\Big) \right)^2 \Bigg)
\Tr \Big(  \phi_5^2+\phi_6^2+\phi_7^2
+\phi_8^2 \Big)  \nonumber \\
 && +\left(\partial_1\partial_2 \left(\frac{1}{e^2}\right)
 -e^2\partial_1\left(\frac{1}{e^2}\right)
 \partial_2\left(\frac{1}{e^2}\right)\right)
 \Tr \Big( (a_2-b_1)\phi_5\phi_6+(a_1-b_1)\phi_7\phi_8 \Big) \nonumber \\
  && +\left(\partial_2\partial_3\left(\frac{1}{e^2}\right) -
  e^2\partial_2\left(\frac{1}{e^2}\right) \partial_3\left(\frac{1}{e^2}\right)\right)
 \Tr \Big( (b_2-c_1)\phi_6\phi_7 +(b_1-c_1)\phi_5\phi_8  \Big) \nonumber \\
 && +\left(\partial_3\partial_1\left(\frac{1}{e^2}\right)
 -e^2\partial_3\left(\frac{1}{e^2}\right) \partial_1\left(\frac{1}{e^2}\right)\right)
 \Tr \Big( (c_2-a_1)\phi_7\phi_5+(c_1-a_1)\phi_6\phi_8\Big)  .
\ea
Note that $a_2-b_1=b_2-a_1= (a_2+b_2-a_1-b_1)/2$ and $a_1-b_1=(a_1-b_1-a_2+b_2)/2$.
The last three terms vanish if $e^2(x,y,z)$ has the factorizable spatial dependency.
When $a_1=a_2=b_1=b_2= c_1=c_2=1/2$, we have $SO(4)$ symmetry which rotates
$\phi_{5,6,7,8}$.  If the coupling constant depends only on the radial variable
 $e^2(\sqrt{x^{2}+y^{2}+z^{2}})$,
there will be a spatial rotational symmetry. Our analysis on the constraint on the
spinor is the most general and so our Lagrangian is the most general Lagrangian
in 3-dim case.

\noindent{\bf \large Acknowledgment}
We would like to thank Dongsu Bak,  Sungjay Lee, P. Ho, Sumit Das
 for interesting discussions.
This work is supported in part by Seoul Science Fellowship (EK), 
the KOSEF SRC Program through CQUeST at Sogang University
(CK,KML),   KRF Grants No. KRF-2005-070-C00030 (KML), and
the KRF National Scholar program (KML).

\noindent\appendix {Appendix: \bf Moving Janus}

Let us consider the space-time dependent coupling constant
$e^2(x+t)$. The supersymmetric condition on the constant spinor
parameter is
\be \Gamma^{01}\epsilon=\epsilon \label{lightcone}. \nonumber \ee
Under the infinitesimal supersymmetric transformation, the
original Lagrangian transforms as in Eq.~(\ref{d0l0}). 
Since $\partial_0 e^2=\partial_1 e^2$ and $\bar{\epsilon}
(\Gamma^0+\Gamma^1)=0$, the Lagrangian is invariant under 1/2 of
16 supersymmetries satisfying the above  condition. We
can mix this time-dependent Janus with other Janus, preserving
some supersymmetry if the supersymmetry conditions are
compatible.

\end{document}